\documentclass[10pt, conference]{IEEEtran}
\IEEEoverridecommandlockouts
\usepackage[left=0.6in,right=0.6in,bottom=0.98in, top=0.63in]{geometry}

\usepackage{cite}
\usepackage{multicol}

\usepackage{amssymb}

\usepackage{amsmath}
\usepackage{amsfonts}
\usepackage[T1]{fontenc}
\usepackage{algpseudocode}
\usepackage[utf8]{inputenc}
\usepackage[english]{babel}
\usepackage[usenames, dvipsnames]{color}
\usepackage{pifont}
\usepackage{booktabs}
\usepackage{mathtools}
\usepackage{textcomp}
 \usepackage{graphicx}
\usepackage{subfigure}
\usepackage[ruled,vlined,commentsnumbered]{algorithm2e}
\usepackage{amsthm}
\usepackage{setspace}

\usepackage{fancyhdr}
\fancypagestyle{firstpage}{
  \lhead{This work has been accepted for publication in the IEEE ICC 2026 Conference.}
}

\SetKwInput{KwIn}{Inputs}

\theoremstyle{definition}

\theoremstyle{remark}

\theoremstyle{plain}

\newcommand{\RNum}[1]{\uppercase\expandafter{\romannumeral #1\relax}}

\def\BibTeX{{\rm B\kern-.05em{\sc i\kern-.025em b}\kern-.08em
    T\kern-.1667em\lower.7ex\hbox{E}\kern-.125em}}

\begin{document}

\title{Outlier-Resistant Fusion for Multi-static Positioning using 5G NR Signals}

\author{
Maximiliano Rivera Figueroa$^{*}$, Jannis Held$^{*}$, Pradyumna Kumar Bishoyi$^{\dagger}$, and Marina Petrova$^{*}$\\[2pt]
$^{*}$Chair of Mobile Communications and Computing (MCC), RWTH Aachen University, Aachen, Germany\\
$^{\dagger}$Department of Electrical Engineering, Indian Institute of Technology Jodhpur, Rajasthan, India\\
Emails: \{maximiliano.rivera@mcc., jannis.held@, petrova@mcc.\}rwth-aachen.de, pradyumna@iitj.ac.in
}

\maketitle
\thispagestyle{firstpage}
\begin{abstract}

Indoor positioning faces ongoing challenges due to complex propagation conditions, such as multipath propagation, signal blockages, and intrinsic target characteristics that substantially impact measurement reliability and positioning accuracy.
Existing methods, in particular Least Squares (LS), frequently struggle to maintain robustness when confronted with unreliable observations caused by multipath interactions and extended targets.
In this work, we propose an outlier-resistant algorithm designed to mitigate the impact of outlier measurements and accurately estimate the position of an extended target in multipath-rich environments. We develop a two-step algorithm in which an initial coarse position estimate is obtained using the angle-of-arrival (AoA) and subsequently refined using the Cauchy loss function to suppress outliers.
The numerical results confirm that the proposed algorithm improves robustness and accuracy, outperforming existing benchmark methods, such as Iterative Reweighted Least Squares (IRLS), LS, and Huber loss function, and achieving a positioning error of less than $70$ cm in $90\%$ of cases. Its effectiveness in mitigating multipath effects is further assessed by comparing tracking performance in cluttered and empty room scenarios.

\end{abstract}
\begin{IEEEkeywords}
Multi-static, Positioning, Frequency Range 2 (FR2), Passive sensing, Robust estimation, Tracking
\end{IEEEkeywords}

\section{Introduction} \label{sec:Introduction}
Next-generation applications, such as smart industries, smart homes, and digital twins, are poised to demand even stricter quality of service (QoS) requirements from beyond 5G networks, i.e. 6G. These mission-critical environments require ultra-precise indoor positioning to coordinate automated robots or track humans for automation, safety, and efficient operations \cite{NGMN6G2022}. To meet the stringent positioning demands of these applications, integrated sensing and communication (ISAC) is emerging as a key technology enabler in 6G systems design \cite{wild20236g,figueroa2024JCS}. ISAC enables the utilization of wireless infrastructure for both communication and sensing by reusing the same hardware, firmware, software, and spectral resources, where
the latest 3GPP release \cite{Rel19.22.837} demonstrate the potential of ISAC for sensing applications.

Various architectures, primarily monostatic and bistatic systems, have been investigated for passive target localization. Monostatic systems can perform coherent processing compared to their counterparts, but suffer from self-interference, which can saturate the receiver if not properly managed. 
In bistatic systems, synchronizing the transmitter and the receiver is challenging; however, they do not face the problem of self-interference, making them an attractive system to deploy. 
Combining these systems yields a multi-static setup, where multiple transmitters and receivers can operate jointly. This configuration improves positioning accuracy by leveraging spatial diversity and aggregating estimates from multiple receivers~\cite{WillisBistaticRadar}.
Despite these advantages, achieving high position accuracy in an indoor environment remains a formidable challenge. Factors such as severe multipath propagation conditions, non-line-of-sight (NLOS) blockages, and dynamic clutter frequently induce outliers in range measurements, degrading the overall positioning accuracy.

Traditional position estimation techniques, such as the least squares (LS) method, are highly sensitive to outlier measurements \cite{Xiong2023Oulier}. These techniques are often built without considering outlier data, leading to non-robust algorithms, where a slight deviation of the data degrades the algorithm's performance. The LS method inherently assumes Gaussian distributed error noise. However, outliers in indoor range measurements follow non-Gaussian, heavy-tailed distributions caused by transient NLOS conditions and multipath components \cite{Gustafsson_TSP}. As a result, even a small number of corrupted measurements can lead to large estimation biases, resulting in significant errors. 

Several recent works have proposed algorithms to address issues related to indoor positioning. In \cite{MuthineniIRLS2024}, the authors propose to use the Iterative Reweighted Least Squares (IRLS) algorithm to cope with outliers that arise from multipath components in the time-difference-of-arrival (TDoA)  measurements. In \cite{Xiong2023Oulier}, different robust loss functions, such as the smoothed $\ell_1$ 
and the Cauchy loss function, are evaluated for time-of-arrival (ToA) measurement errors using a two-component bivariate Gaussian mixture distribution to account for anomalous errors. 

\textit{While these studies show promising results, they overlook scenarios where multipath components interact with passive targets, a common real-world condition that increases the likelihood of outliers in measurements. This highlights the need for robust fusion strategies that can handle unreliable observations.}


Beyond environmental uncertainties, the target’s spatial extent introduces additional challenges for indoor positioning. Extended targets, such as humans or robots, reflect signals from multiple scatter points, biasing ToA estimates toward the closest reflection rather than the true centroid~\cite{marus_extended}. This results in range biases and heavy-tailed, non-Gaussian errors that degrade positioning accuracy.

To address the above-mentioned issues, in this paper, we present \textbf{an outlier-resistant algorithm} that is able to withstand the impact of unreliable observations while fusing the estimates to find the position of an extended passive target in a real-world indoor scenario. 
We assume a 5G New Radio (NR) compliant multi-static system in an indoor environment that operates in the FR2, consisting of one transmitter and multiple receivers.
The ToA estimates of the receivers are fused to estimate the position of the extended passive target. In order to make the estimation robust against outliers we employ the Cauchy loss function in the fusion step, thereby reducing the influence of large errors in the measurements. Thus, the proposed algorithm aims at suppressing the impact of the outliers arising from the multipath components and the intrinsic characteristics of the extended target. In contrast to the IRLS algorithm, the proposed outlier-resistant method does not require assigning weights to individual estimates, as it intrinsically mitigates the influence of errors through the inherent properties of the function employed in the fusion process.
The main contributions of our work are
\begin{itemize}
\item We present a novel two-step algorithm that accurately estimates the position of an extended passive target in a realistic indoor environment, accounting for challenges introduced by multipath propagation and target characteristics. The first step provides a coarse estimate using AoA measurements, followed by refinement via gradient descent (GD) over an outlier-resistant fusion method. The performance evaluation indicates that our algorithm outperforms different benchmarks, such as IRLS and LS, in terms of positioning error.
\item  We develop an outlier-resistant fusion method that incorporates the Cauchy loss function to inherently suppress the influence of outliers without requiring the assignment of explicit weights to individual estimates. 
\item We conduct a comprehensive evaluation using a realistic simulation framework based on a quasi-deterministic raytracer \cite{LecciQD} and an extended passive target modelled as a human-shaped object with 17 reflection points. The effectiveness of the proposed algorithm is demonstrated through extensive comparisons with benchmark methods. Furthermore, we provide insights into the impact of obstacles on positioning performance in multipath-rich environments by comparing cluttered and empty room scenarios.
\end{itemize}
The remainder of the paper is organized as follows. Section \ref{sec:SystemModel} introduces the system model. In Section \ref{sec:PowerAllocation}, the positioning problem is presented, followed by the proposed algorithm in Section \ref{sec:Proposedsoln}. 
Subsequently, Section \ref{sec:PerformanceEvaluation} discusses the performance evaluation, followed by conclusions in Section \ref{sec:Conclusions}.
\section{System Model}\label{sec:SystemModel}
We consider an indoor scenario composed of a group of $\mathcal{K} = \{1, 2, \cdots, K\}$ gNodeBs (gNBs) deployed in an indoor environment. The gNBs have sensing capabilities enabling the establishment of bistatic links among themselves. We assume that all gNBs are synchronized\footnote{The effect of synchronization on the accuracy is out of scope for this study.} and coordinated by a central unit (CU). Each gNB is deployed with a uniform linear array (ULA) and follows the 5G NR standard operating in FR2. The system aims to locate and track an extended target with an unknown position $\mathbf{x}$ that moves freely within the indoor area. Each gNB is at a known position $\mathbf{x}_k \in \mathbb{R}^3$, $\forall k \in \mathcal{K}$. 

To estimate the position of the extended target, the system operates in bistatic sensing mode. In this configuration, one gNB is designated as the transmitter, denoted by $k$ with $k \in \mathcal{K}$, while another gNB, $j \neq k$, acts as the receiver. The transmitter sends an Orthogonal Frequency Division Multiplexing (OFDM) signal, which is reflected by the target and received by the $j$-th gNB. The receiver then estimates the time-of-flight (ToF), i.e., the bistatic distance, along with the angle-of-arrival (AoA). This setup can be extended to include multiple receivers, forming a multi-static sensing configuration~\cite{WillisBistaticRadar}, where the signal transmitted by the $k$-th gNB is collected by all $j$-th gNBs, with $j \in \mathcal{K} \setminus {k}$.
The transmitted signal comprises $N_s$ active subcarriers with a subcarrier spacing of $\Delta_f$ and $N_M$ OFDM symbols. The transmitted baseband signal from gNB $k \in \mathcal{K}$ is
\begin{equation}
s_k(t) = \sum_{m=0}^{N_M-1} \sum_{n=0}^{N_s-1} c^k_{m, n} e^{j\, 2\pi n \Delta_f t} g(t - mT),\end{equation}
where $c_{m, n}^k$ represents the complex modulation symbols transmitted for a given subcarrier $n$ and symbol $m$. The pulse shaper is represented by the function $g(\cdot)$, and $T$ is the overall OFDM symbol duration, consisting of the symbol duration $T_s$ and the cyclic prefix (CP) duration $T_{CP}$. 

Let $\rho_k$ denote the transmit power of the $k$-th gNB with $0 \leq \rho_k \leq P_{\text{max}}$, where $P_{\text{max}}$ is the maximum power that any gNB can transmit, and $\mathbf{w}_{tx, k} \in \mathcal{C}^{N_{tx}\times 1 }$ denotes the transmit precoding vector of the $k$-th gNB, with $N_{tx}$ the total number of antenna elements in the transmitter. Then, the transmit signal from the $k$-th gNB is $\mathbf{b}_k(t) = \sqrt{\rho_k}\,\mathbf{w}_{tx, k} \,s_k(t)$.


The received signal at the $j$-th gNB is expressed as 
\begin{equation}
    \mathbf{y}_{k, j} (t) =  h_{k, j}(t) \mathbf{b}_k(t) + \mathbf{n}_{j}(t), \label{eq:rx_signal}
\end{equation}
where $h_{k, j}(t)$ is the bistatic radar channel between the $j$-th gNB and the $k$-th gNB, which accounts for the effect of the target radar cross-section (RCS) and the signal propagation path-loss over a multipath environment, and $\mathbf{n}_j(t)$ denotes the noise vector. The receiving beamforming $\mathbf{w}_{rx, j}$ is applied to Eq.~\eqref{eq:rx_signal} to combine the signal from each antenna element, given as $y_{k, j}(t) = \mathbf{w}_{rx, j}^T \,\mathbf{y}_{k, j}(t)$.

For a given pair of precoding vectors $(\mathbf{w}_{tx,\,k},\,\mathbf{w}_{rx,\,j})$, the following steps are considered to process the received signal. First, CP removal is performed, followed by the Fourier Transform over each received symbol and applying zero-forcing (ZF) to remove the modulation symbols. Then, the received signal of the $m$-th OFDM symbol and $n$-th subcarrier at the $j$-th gNB $D_{m, n}^{k, j}$, is represented by \cite{figueroa2024WCNC}:
\vspace{-0.2cm}
\begin{equation}
    D_{m, n}^{k, j} = \sum_{\ell = 0}^{L-1} \gamma_{\ell}^{k, j} \,e^{j2\pi\,(m\,T_s f_{D, \ell}^{k, j} - n\tau_{\ell}^{k, j} \Delta f) } + N_{m, n, j},
    \label{eq:DmnMultipath}
\end{equation}
where $\gamma_{\ell}^{k, j}$ is the radar channel gain which accounts for the path loss, transmit power, beamforming gain, and the RCS for the $\ell$-th path, $f_{D, \ell}^{k, j}$ is the Doppler shift seen by the $j$-th gNB over the $\ell$-th path, and $\tau_{\ell}^{k, j}$ is the propagation time delay of the $\ell$-th path from the $k$-th gNB to the $j$-th gNB. The noise of the $m$-th OFDM symbol and $n$-th subcarrier at the $j$-th gNB is represented as $N_{m, n, j}$. $L$ represents the total number of paths.

To estimate the bistatic distance, first, we use the Periodogram technique given by
\begin{equation}
A_{k, j}(u, v) = \Bigg| \sum_{p=0}^{N_s^{'}-1} \Bigg(\sum_{q=0}^{N_M'-1} D_{p, q}^{k, j} e^{-2\pi j q\,v/N_M'} \Bigg)  e^{2\pi j q\,u/N_s'} \Bigg ) \Bigg|^2, \label{eq:Periodogram}
\end{equation}
where $N_s' \geq N_s$ and $N_M' \geq N_M$. $A_{k, j}(u, v)$ is the Periodogram amplitude for the link from the $k$-th gNB to the $j$-th gNB, and the indexes $u$ and $v$ represent the distance and radial velocity bins.
Then, the bistatic distance is computed as 
\begin{equation}
    \hat{d}_{k, j} = \frac{\hat u c_0}{\Delta f\, N_{s}'}, \label{eq:distanceEstimation}
\end{equation}
with $\hat u$ being the index that maximizes Eq.~\eqref{eq:Periodogram}.
Therefore, we define a set of estimated bistatic distances as $\{\hat{d}_{k, j} \}_{j=1, j\neq k}^{K}$ that defines the multi-static system, with one transmitter and $K-1$ receivers. 
By fusing these bistatic distances, the target's position is estimated. This fusion process must account for the varying error levels associated with each estimate to ensure reliable positioning.
To improve the estimation accuracy of Eq.~\eqref{eq:distanceEstimation}, we follow the clutter removal approach presented in \cite{Henninger2023Crap1}, where measurements are taken under target-free conditions.

\vspace{-0.17cm}
\section{Position Estimation in Multi-static Systems}\label{sec:PowerAllocation}

Estimating the position of a target based on the set of measurements $\{\hat{d}_{k, j}\}_{j=1, j\neq k}^{K}$ becomes challenging when the error distribution is unknown. Multipath effects often introduce biases in the distance estimates, causing deviations from the true values~\cite{FigueroaMCIS}. Additionally, as noted in~\cite{WillisBistaticRadar}, the radar cross section (RCS) observed by different receivers can vary significantly, especially for extended targets. This variability leads to unequal error levels across measurements, as also discussed in~\cite{Karl2016EOT}, and may result in outliers that degrade the accuracy of the final position estimate. Moreover, when aggregating data from multiple gNBs, the heterogeneity in error characteristics across estimates introduces uncertainty into the fusion process, as no prior information is available regarding their reliability.

A common approach for estimating the position of the target involves the formulation of a minimum LS problem, which fuses the set of measurements $\{\hat{d}_{k, j}\}_{j=1, j\neq k}^{K}$, for a given transmitter $k$, is:
\begin{equation}
\min_{\mathbf{x}} \sum_{j = 1, \,j\neq k}^K \left(\hat{d}_{k, j} - \|\mathbf{x}_k - \mathbf{x}\| - \|\mathbf{x} - \mathbf{x}_j\| \right)^2. \tag{P1}
\label{problem:P1}
\end{equation}

By identifying the argument that minimizes \ref{problem:P1}, the position of the target can be estimated. 
However, this approach assumes that the noise is identically distributed across all measurements, which is typically not the case in these scenarios, as previously discussed. This issue becomes even more pronounced when all possible transmitters are considered, i.e., varying $k \in \mathcal{K}$. In this case, the problem is reformulated as:
\begin{equation}
\min_{\mathbf{x}} \,\sum_{k = 1}^K\,\,\sum_{j = 1, \,j\neq k}^K \left(\hat{d}_{k, j} - \|\mathbf{x}_k - \mathbf{x}\| - \|\mathbf{x} - \mathbf{x}_j\| \right)^2. \tag{P2}
\label{problem:P2}
\end{equation}

Rotating the selection of the transmitter can be interpreted as a round-robin (RR) approach, where only one gNB is used as the transmitter at a time.

The LS formulation in \ref{problem:P1} and \ref{problem:P2} is known to perform poorly with outlier measurement data, since the square loss amplifies the error regardless of its magnitude \cite{Zoubir2012RobustMag, PENNACCHI2008923}. Moreover, this approach assumes that the noise level is identically distributed across the collected measurements, which is not commonly found in these scenarios. Therefore, we propose the following outlier-resistant approach to cope with outlier measurements and non-identical noise distribution within the measurements.

Inspired by \cite{Xiong2023Oulier}, we modify the fusion scheme of \ref{problem:P2} by replacing the square loss function with the Cauchy loss function. The Cauchy loss function is defined as:
\begin{equation}
\psi_{\eta} (z) =  \frac{\eta^2}{2} \log\left(1 + \frac{z^2}{\eta^2} \right),
\label{eq:CauchyLoss}
\end{equation}
where $\eta$ is a penalty factor. A larger value of $\eta$ decreases the significance assigned to the error $z$, and its value determines the robustness of the outlier-resistant approach \cite{Xiong2023Oulier}.

The Cauchy loss function can reduce the impact of outlier measurements compared to the square loss function by assigning less importance to larger errors. This function is robust in scenarios with non-Gaussian noise distributions, making it particularly useful for environments where measurement errors are not identically distributed. Hence, \ref{problem:P2} can be rewritten as:
\begin{equation}
\min_{\mathbf{x}} \,\sum_{k = 1}^K\,\sum_{j = 1, \,j\neq k}^K \psi_{\eta}\left(\hat{d}_{k, j} - \|\mathbf{x}_k - \mathbf{x}\| - \|\mathbf{x} - \mathbf{x}_j\| \right) \tag{P3}.
\label{problem:P3}
\end{equation}

Solving \eqref{problem:P3} is challenging because the problem is not necessarily convex, and hence convergence to a global minimum cannot be guaranteed. 
Consequently, the initial guess must be sufficiently accurate to ensure convergence to the correct minimum.
\vspace{-0.1cm}
\section{Proposed Solution}\label{sec:Proposedsoln}
\vspace{-0.1cm}
In this section, we present the proposed algorithm for estimating the target position in the presence of outlier data. 
\subsection{Finding the Optimal Position}
To solve \ref{problem:P3}, GD is employed to identify a local minimum. Therefore, providing an accurate initial guess is required to ensure convergence to the desirable local minimum. 
We utilize the available AoA estimation to find an initial position estimate.

As outlined in \cite{FigueroaMCIS}, the target's position can be determined using the AoA estimates through the maximum likelihood (ML) method. This approach uses the VMF distribution as the likelihood function, similar to the Gaussian distribution but defined in a two-dimensional spherical space. The ML method is given by
\begin{equation}
\hat{\mathbf{x}}_{\text{AoA}} = \arg\max \, \prod_{k = 1}^K \,\, \prod_{j = 1, j \neq k}^K q_{k, j}^{\text{AoA}}, \label{eq:AoAML}
\end{equation}
where $q_{k, j}^{\text{AoA}}$ is the likelihood function of the VMF distribution when the $k$-th gNB is transmitting and the $j$-th gNB is receiving. It is defined as:
\begin{equation}
q_{k, j}^{\text{AoA}} = \frac{1}{2\pi I_0(\kappa_{k,j})} \exp\Big(\kappa_{k,j} \hat{\boldsymbol{u}}_{k, j} \boldsymbol{u}_{k, j}^T\Big),
\end{equation}
where $\boldsymbol{u}_{k, j}$ represents the unit vector that points to the true position of the target from the $j$-th gNB when the $k$-th gNB acts as the transmitter, while $\hat{\boldsymbol{u}}_{k, j}$ denotes the unit vector corresponding to the estimated AoA. The term $I_0(\cdot)$ refers to the modified Bessel function of the first kind and of order zero, and $\kappa_{k,j}$ is the concentration parameter, which is conceptually similar to the standard deviation, for the gNB pair $k$ and $j$. It is important to note that the obtained position is a coarse estimation used to initialise the GD.

Once the initial position is determined, the GD algorithm is utilized to solve the optimization problem \ref{problem:P3}. The iterative process is expressed as:
\begin{equation}
\hat{\mathbf{x}}^{(i+1)} = \hat{\mathbf{x}}^{(i)} - \alpha \nabla_{\mathbf{x}}F(\mathbf{x})\Big|_{\mathbf{x} = \hat{\mathbf{x}}^{(i)}},\label{eq:GradientDescentStep}
\end{equation}
where $(i)$ represents the iteration index, and $\alpha$ denotes the step size. The algorithm proceeds iteratively until the convergence criterion $||\hat{\mathbf{x}}^{(i)} - \hat{\mathbf{x}}^{(i-1)} || \leq \epsilon$ is met, with $\epsilon$ being the predefined convergence threshold. The objective function $F(\cdot)$ is
\begin{equation}
F(\hat{\mathbf{x}}^{(i)}) = \sum_{k = 1}^K \,\,\sum_{j = 1, \, j \neq k}^K f_{\eta,\,k,\,j}(\hat{\mathbf{x}}^{(i)}), \label{eq:ObjectiveFunctionGD}
\end{equation}
where the term $f_{\eta,\,k,\,j}(\hat{\mathbf{x}}^{(i)})$ is expressed as:
\begin{align}
f_{\eta,\,k,\,j}(\hat{\mathbf{x}}^{(i)}) \triangleq  \nonumber \psi_{\eta}\left(\hat{d}_{k, j} - \|\mathbf{x}_k - \hat{\mathbf{x}}^{(i)}\| - \|\hat{\mathbf{x}}^{(i)} - \mathbf{x}_j\| \right).
\end{align}

Then, the gradients are derived by calculating the partial derivatives. For the x-axis, the gradient is given by:
\begin{align}
\frac{\partial f_{\eta,\,k,\,j}(\mathbf{x})}{\partial x}\Bigg|_{\mathbf{x} =  \hat{\mathbf{x}}^{(i)}} =& \,\dfrac{2\delta_{k,j}(\hat{\mathbf{x}}^{(i)}) }{\eta (1 + \delta_{k,j}(\hat{\mathbf{x}}^{(i)})^2/\eta^2)} \times \nonumber \\
&\Bigg( \frac{(\hat{x}^{(i)} - x_j)}{||\hat{\mathbf{x}}^{(i)} - \mathbf{x}_j||} + \frac{(\hat{x}^{(i)} - x_k)}{||\hat{\mathbf{x}}^{(i)} - \mathbf{x}_k||}\Bigg),\label{eq:xGradient}
\end{align}
where $\delta_{k,j}(\hat{\mathbf{x}}^{(i)})$ is defined as:
\begin{equation}
\delta_{k,j}(\hat{\mathbf{x}}^{(i)}) = \hat{d}_{k, j} - \|\hat{\mathbf{x}}_k - \mathbf{x}^{(i)}\| - \|\hat{\mathbf{x}}^{(i)} - \mathbf{x}_j\|.
\end{equation}
The derivation for other axes follows a similar approach. The summary of the algorithm is shown in Algo.~\ref{algo:ALGO1}, where $\mathbf{D}_K = \big\{\{\hat{d}_{k, j}\}_{j=1, j\neq k}^{K}\big\}_{k=1}^K$ is the set of all measurements of bistatic distances, $\mathbf{\Theta}_K=\big\{\{\hat{\theta}_{k, j}\}_{j=1, j\neq k}^{K}\big\}_{k=1}^K$ is the set of all AoA measurements, with $\hat{\theta}_{k, j}$ the AoA obtained when the $k$-th gNB transmits and the $j$-th gNB receives, and $\mathbf{\varkappa}_K = \big\{\{\kappa_{k, j}\}_{j=1, j\neq k}^{K}\big\}_{k=1}^K$ is the set of the concentration parameters.

\begin{algorithm}[t]
	\small
	\SetKwInOut{Input}{Inputs}
	\SetKwInOut{Output}{Output}
    \SetKwInOut{Return}{Return}
	\Input{$\mathbf{D}_K$, $\mathbf{\Theta}_K$, $\mathbf{\varkappa}_K$, $\eta$, $\alpha$, $\epsilon$}
	\Output{ $\hat{\mathbf{x}}^{*}$ }
    Build $q_{k, j}^{\text{AoA}}\, \forall\, k, j \in \mathcal{K}, \, k\neq j$ \\
    Find $\hat{\mathbf{x}}_{\text{AoA}}$ by solving Eq.~\eqref{eq:AoAML}\\
	Set $ converge=0, i=0$, and $\mathbf{x}^{(i)} \leftarrow \hat{\mathbf{x}}_{\text{AoA}}$\\
	\While{converge = 0}{
        Compute the gradients by using Eq.~\eqref{eq:xGradient}\\
        Find $\hat{\mathbf{x}}^{(i+1)}$ with Eq.~\eqref{eq:GradientDescentStep}\\
        $i \leftarrow i+1$\\
		\uIf{$||\hat{\mathbf{x}}^{(i)} - \hat{\mathbf{x}}^{(i-1)}||<\epsilon$}
        { $\hat{\mathbf{x}}^{*} \leftarrow \hat{\mathbf{x}}^{(i)}$ \\
        $converge \leftarrow 1$}
	}
    \Return{$\hat{\mathbf{x}}^{*}$}
	\caption{Two-Step Outlier-Resistant 
    Algorithm}
	\label{algo:ALGO1}
\end{algorithm}

\subsection{Complexity Analysis}
Let $Y$ denote the number of iterations required for the GD algorithm to converge. The computational complexity of the GD approach is $\mathcal{O}(Y\,d\,\nu)$, where $d$ represents the dimensionality of $\mathbf{x}$, and $\nu$ is the number of estimations used in the fusion process of Eq.~\eqref{eq:ObjectiveFunctionGD}. By setting $\nu = K \cdot (K-1)$ and noting that solving Eq.~\eqref{eq:AoAML} involves a coarse estimation with lower complexity compared to the GD algorithm, the overall complexity of the proposed Algo.~\ref{algo:ALGO1} is $\mathcal{O}(Y  d  K^2)$ \cite{boydconvex}.

\vspace{-0.05cm}
\section{Performance Evaluation}\label{sec:PerformanceEvaluation}

%
\subsection{Setup and Scenarios}\label{sec:SetupScenario}



\begin{figure}[t]
\vspace{-0.5cm}
\centering
	\subfigure[]{
		\includegraphics[trim={0cm 0cm 0 0cm},clip, width=0.65\linewidth]{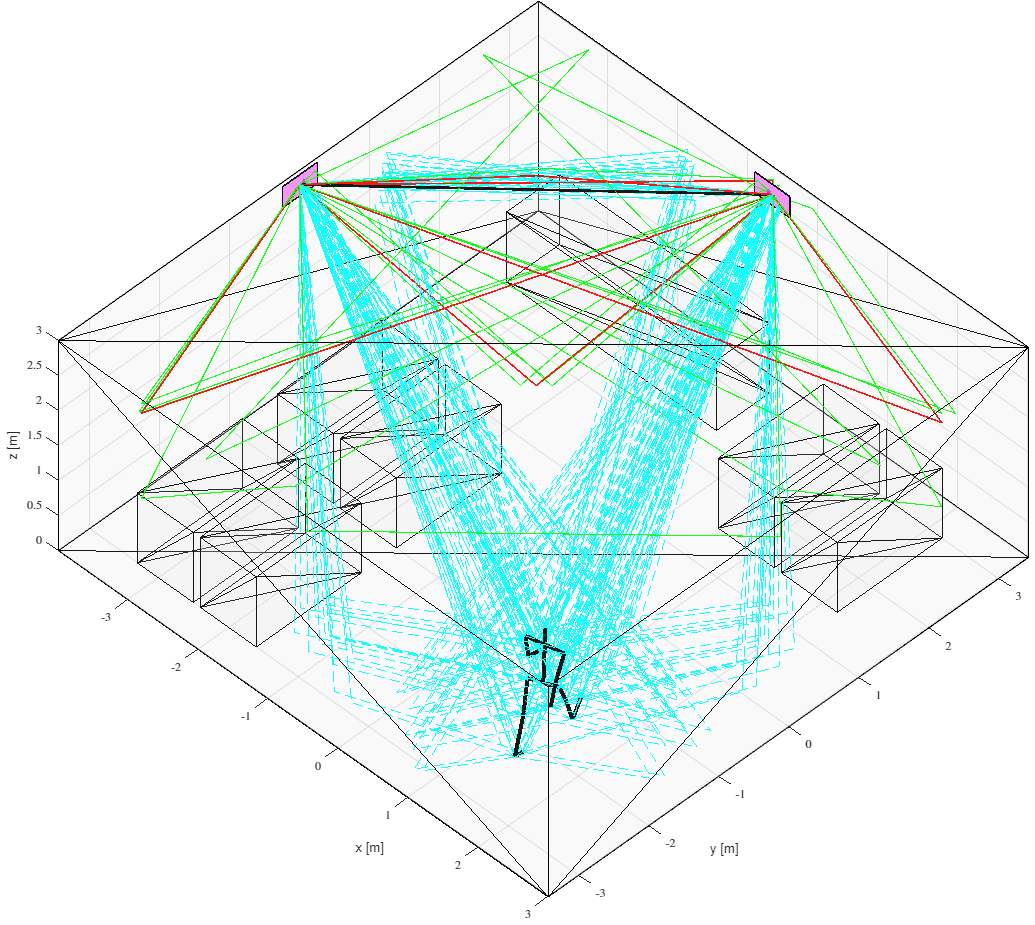}
    \label{fig:QD}}

	\subfigure[]{    \includegraphics[angle = -90, trim={3cm 3cm 1.5cm 3cm},clip, width=0.75\linewidth]{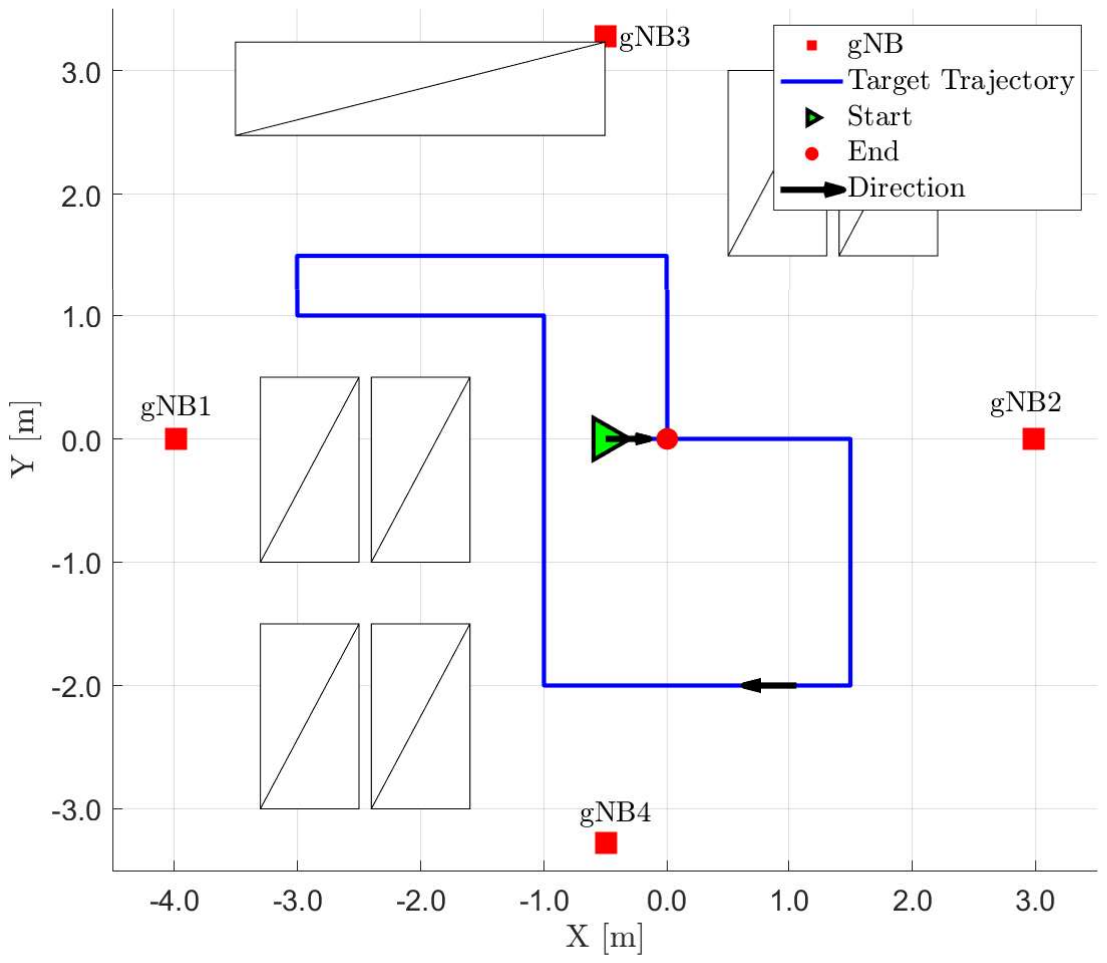}
    \label{fig:Scenario}
	}
	\caption{From top to bottom: (a) QD raytracer illustration, (b) Indoor scenario. The blue line is the path of the target; the red squares indicate the gNBs, and the crossed rectangles represent objects placed in the room.}
	\label{fig:Distance_K2}
    \vspace{-0.5cm}
\end{figure}

To evaluate the proposed outlier-resistant fusion algorithm, we developed a framework incorporating a realistic raytracing model and an extended target model with MATLAB. 
The raytracer uses a quasi-deterministic (QD) channel model and an extended target model based on a human walking profile represented by 17 reflection points~\cite{LecciQD}.

Fig.~\ref{fig:QD} shows the QD raytracer and the extended target in an indoor common room scenario, where the target moves around the space. Various objects have been strategically placed to create obstacles, adding to the complexity of the environment. This figure exclusively depicts two gNBs configured in a bistatic setup within a multipath scenario, where the maximum number of path reflections is set to two.

Fig.~\ref{fig:Scenario} depicts the simulated scenario where four gNBs ($K = 4$) are deployed. The figure shows the locations of the obstacles, the path followed by the target, and the direction of movement. Each gNB has a ULA consisting of four antenna elements, modelled using MATLAB's Antenna Toolbox \cite{MATLAB}. The system operates at $28$ GHz of carrier frequency with $400$ MHz of bandwidth. We set the transmit power to $23$ dBm.

The simulation is conducted as follows. The target is positioned at the next location along the path at each time step. For each target position, a round-robin (RR) mechanism is used, with one gNB as the transmitter and three as receivers. We employ an omnidirectional beam for the transmitter precoding vector, and we apply the Multiple Signal Classification (MUSIC) algorithm to estimate the AoA, which is subsequently used for the receiver's precoding vector. Once the AoA and bistatic distances are estimated, the proposed Algo.~\ref{algo:ALGO1} is executed to calculate the target's position. 
We set $\eta = 2.3849\sigma$, where $\sigma$ denotes the standard deviation of the measurement noise and it is computed following~\cite{MuthineniIRLS2024}, as it attains approximately $95\%$ of the asymptotic efficiency under the assumption of Gaussian noise~\cite{PENNACCHI2008923}.
Finally, using the estimated position, we employ a standard Kalman Filter (KF) to track the target over the path. 

We assess the performance of the proposed algorithm by comparing it with several well-known loss functions, such as the $\ell_2$ loss function (LS problem formulation) and Huber loss function, and to the IRLS algorithm, which is a robust technique to overcome the presence of outliers. The parameter used for the Huber loss function is obtained from \cite{PENNACCHI2008923}.

\subsection{Simulation Results and Discussion}\label{sec:SimResults}

\begin{figure}[t]
\centering
	\subfigure[]{
		\includegraphics[trim={0cm 6.5cm 0 7.3cm},clip, width=0.72\linewidth]{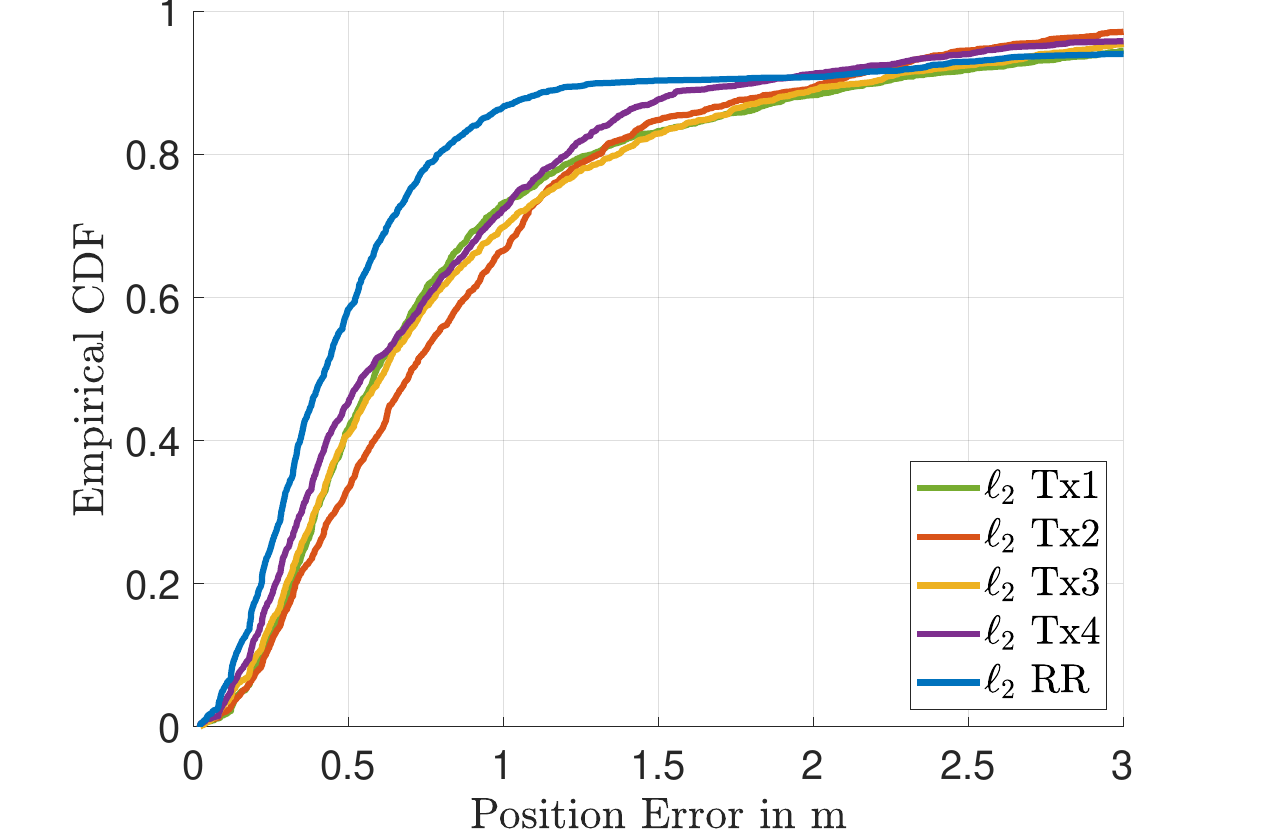}
    \label{fig:TxComparison}}
	\subfigure[]{\includegraphics[trim={0cm 6.5cm 0 7.3cm},clip, width=0.72\linewidth]{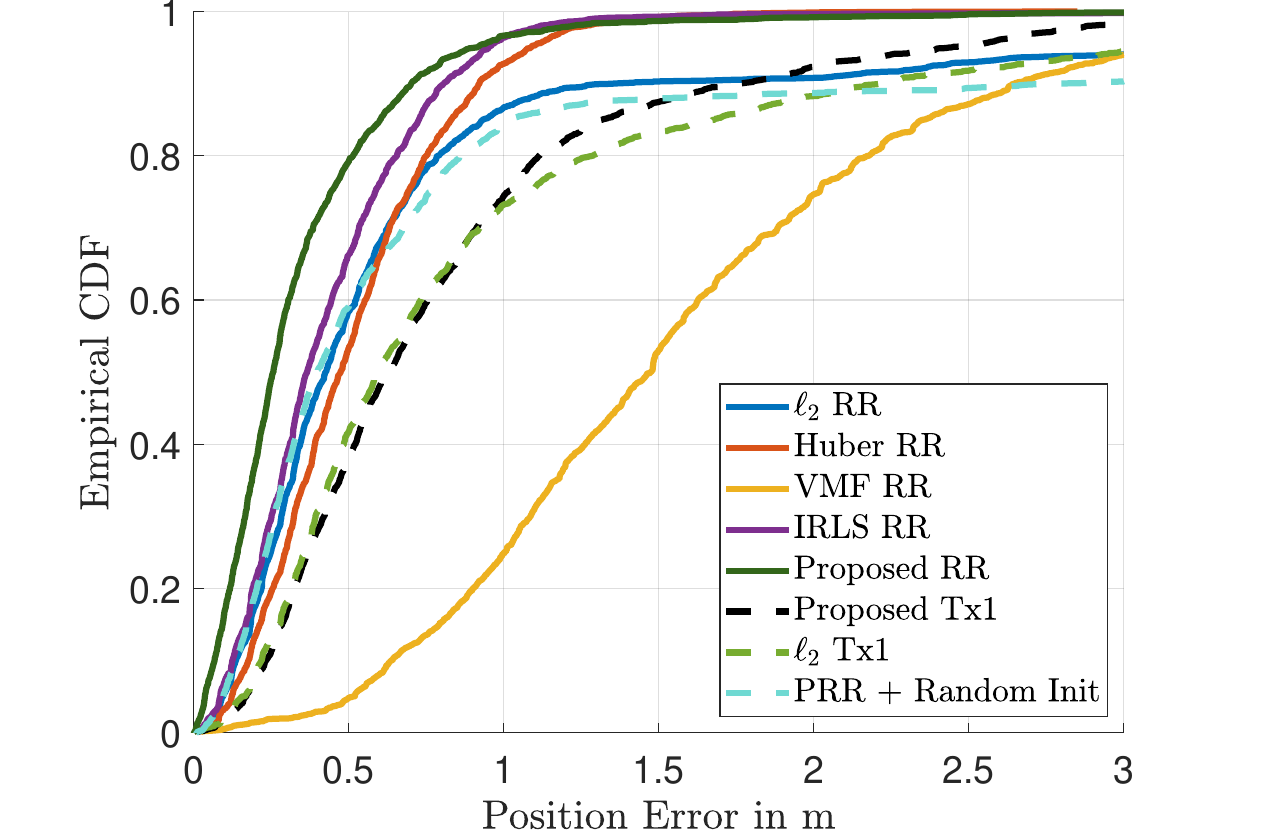}
    \label{fig:Cauchy_vs_benchmarks}
	}
	\caption{From top to bottom: (a) CDF of the position error of \ref{problem:P1}, which uses one transmitter at a time, and \ref{problem:P2}, which employs a round-robin approach to fuse all measurements. (b) Positioning error comparison between the proposed algorithm and benchmarks. PPR: Proposed RR.}
	\label{fig:CDF_Error}
    \vspace{-0.2cm}
\end{figure}


\begin{figure}[t]
    \centering
    \includegraphics[trim={1.6cm 6.8cm 2cm 7.3cm},clip, width=0.75\linewidth]{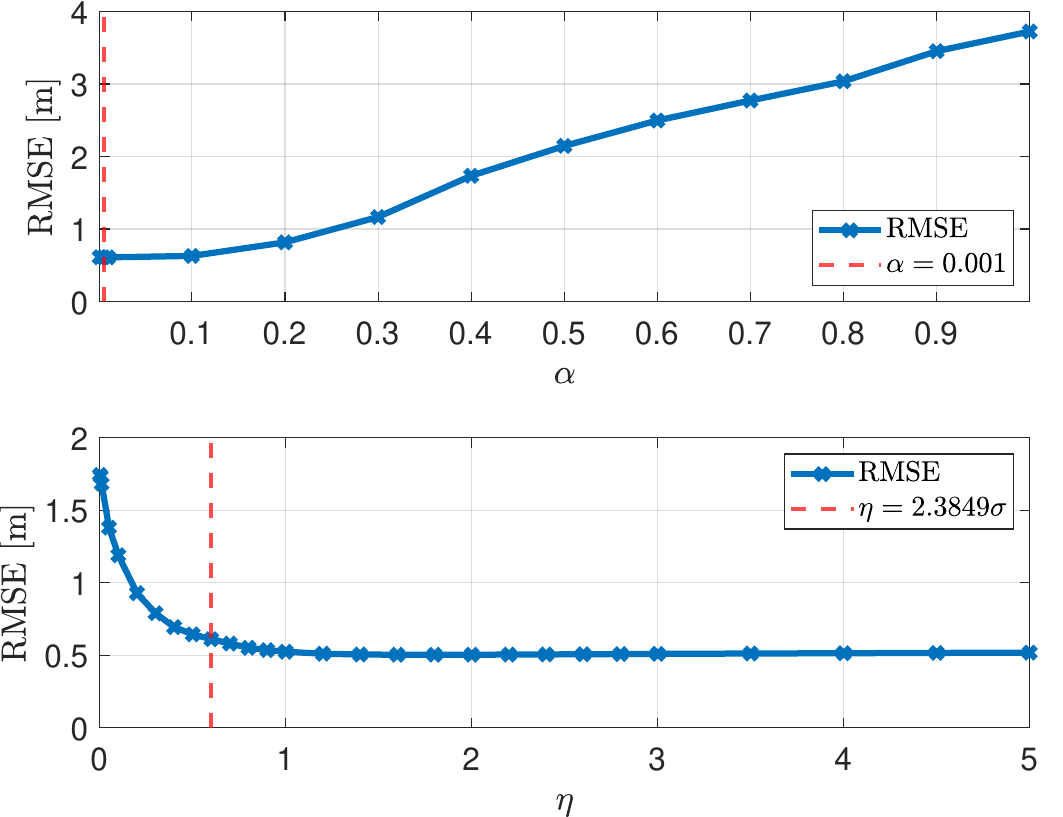}
    \caption{Root mean square error (RMSE) of the position estimation as a function of $\alpha$ (top) and $\eta$ (bottom).}
    \label{fig:alpha-eta}
    \vspace{-0.6cm}
\end{figure}



\begin{figure}[t]
\centering
	\subfigure[]{
		\includegraphics[angle = -90, trim={2cm 3cm 0 3cm},clip, width=0.6\linewidth]{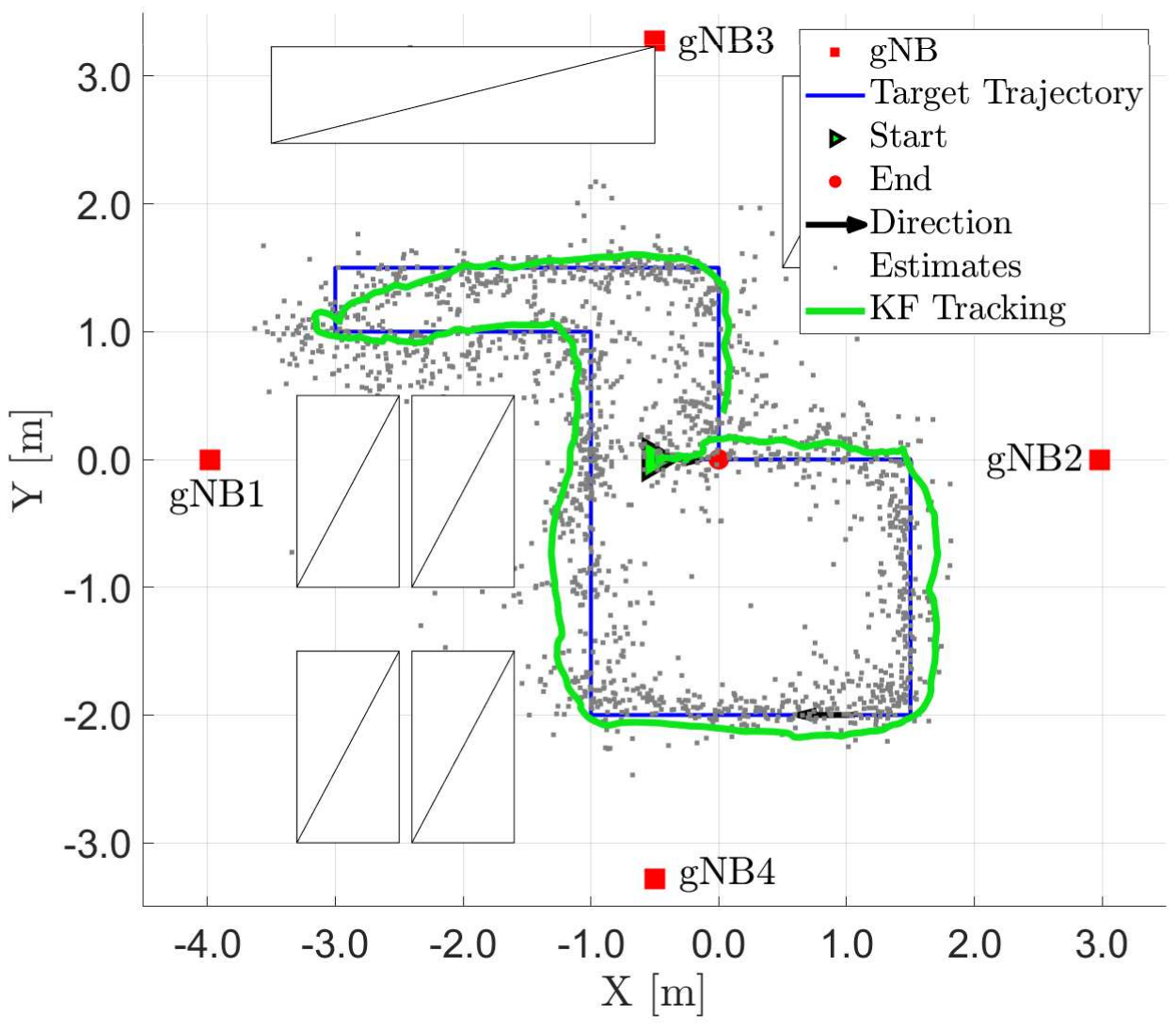}
    \label{fig:KFTrackingFullRoom}}
	\subfigure[]{
    \includegraphics[angle = -90, trim={2cm 3cm 0 3cm},clip, width=0.6\linewidth]{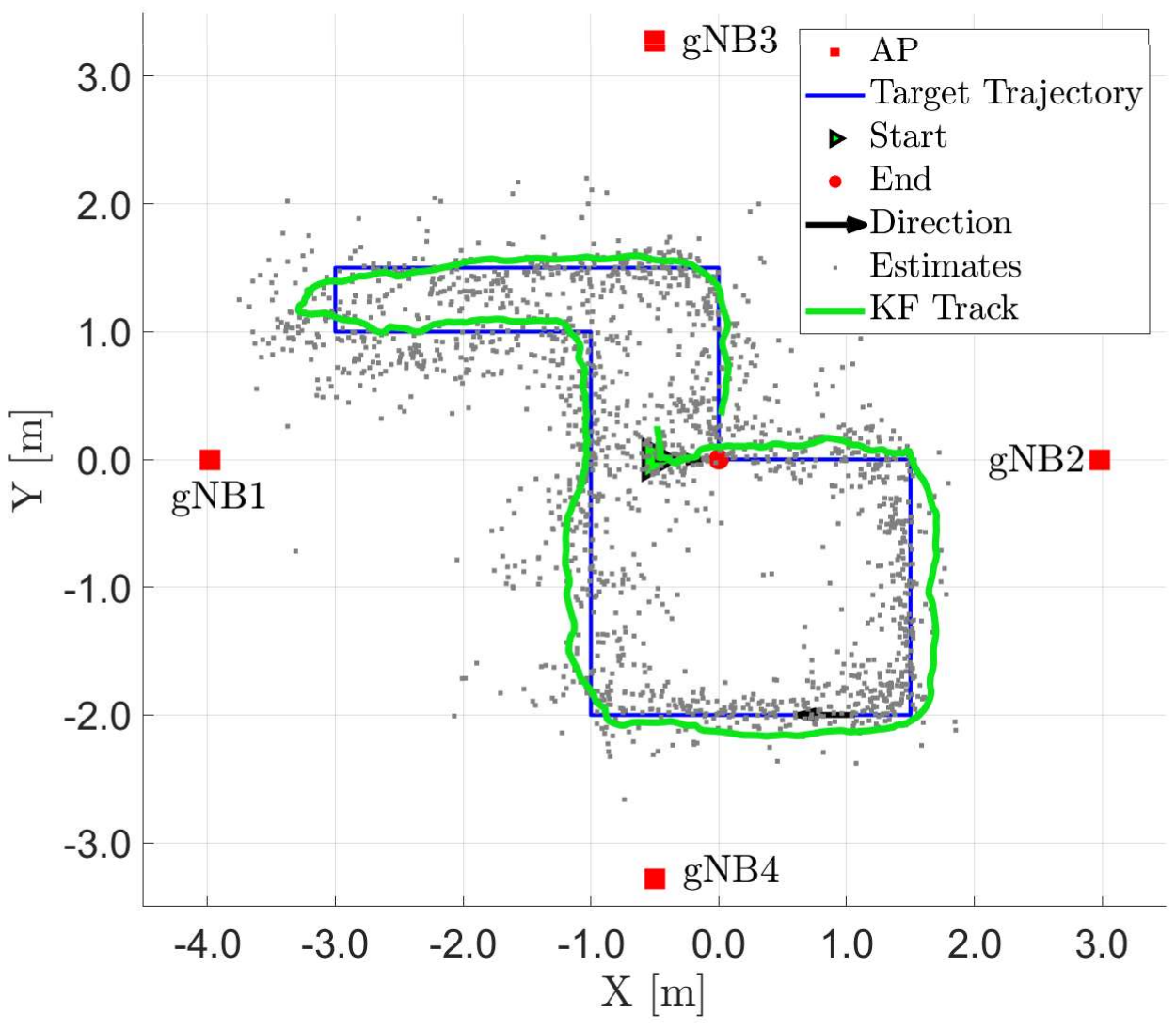}
    \label{fig:KFTrackingEmptyRoom}
	}
	\caption{From top to bottom: (a) Tracking performance on the proposed scenario, (b) Tracking performance for an empty room scenario.}
	\label{fig:KFTracking}
    \vspace{-0.3cm}
\end{figure}
Fig.~\ref{fig:CDF_Error} illustrates the estimator's performance by showing the cumulative density function (CDF) of the position error estimation prior to the tracking stage. Fig.~\ref{fig:TxComparison} compares the estimator's performance when the position is determined using a single gNB, labelled as Tx1 to Tx4, and when the RR approach is employed. It can be clearly seen that the four gNBs behave almost identical, and that combining all the estimation in the RR approach using the least square estimator brings slight gains in the accuracy of the estimation. 

In Fig.~\ref{fig:Cauchy_vs_benchmarks}, the performance of the proposed algorithm is compared with several benchmarks for the RR approach. The results demonstrate that the proposed algorithm outperforms the benchmarks, achieving a positioning accuracy of $70$ cm or less in $90\%$ of the cases.
The findings indicate that using the $\ell_2$ loss function becomes suboptimal as the number of fused estimates increases, due to the presence of outliers. Similarly, the Huber loss function offers no noticeable fusion gain, as it uses the $\ell_2$-norm for the inlier data~\cite{PENNACCHI2008923}. On the other hand, while the VMF distribution yields lower accuracy compared to the other estimators, it serves only as the initialization method for the GD algorithm. 
We also evaluate the performance of the proposed algorithm under random initialization. As shown by the dashed light blue line, random initialization fails to achieve the same level of accuracy, highlighting the importance of a good starting point in the algorithm. Moreover, the effectiveness of the method depends on the number of estimates involved in the fusion process. As more estimates are combined, greater fusion gains are obtained. As illustrated by the dashed black line in the figure, the improvement in positioning accuracy becomes limited when only a small number of estimates are available.

In Fig.~\ref{fig:alpha-eta}, the root mean square error (RMSE) of the position estimation is used to evaluate the impact of the $\alpha$ and $\eta$ parameters. The top plot shows that increasing the GD step size $\alpha$ degrades positioning accuracy, as the algorithm fails to converge to the correct local minimum. The red dashed line indicates the value used in simulations. The bottom plot illustrates the effect of the Cauchy parameter $\eta$ on outlier rejection. Smaller $\eta$ values perform poorly, while larger ones improve robustness and approach asymptotic efficiency. We set $\eta = 2.3849\sigma$, which achieves optimal efficiency under Gaussian noise. However, due to multipath and intrinsic target characteristics, the measured noise in this scenario deviates from the Gaussian model, suggesting that a larger $\eta$ may be needed. A detailed selection of the optimal $\eta$ for such conditions is out of the scope of this study.

Finally, Fig.~\ref{fig:KFTracking} shows the estimated path using a standard Kalman Filter (KF) applied to the position output of the proposed algorithm. Fig.~\ref{fig:KFTrackingFullRoom} presents the tracking results in the room depicted in Fig.~\ref{fig:Scenario}, while Fig.~\ref{fig:KFTrackingEmptyRoom} shows the results for an empty room without obstacles. 
The RMSE of the positioning error is $0.59$ m for the empty room and $0.61$ m for the room with obstacles. This close RMSE values demonstrate the robustness of the proposed outlier-resistant algorithm, which remains accurate even in cluttered environments.
A noticeable difference occurs in the upper-left section of the path, where a U-turn takes place. In the empty room, the estimated path closely follows the ground truth, whereas in the cluttered room, the track deviates at the U-turn. This result highlights the importance of considering environmental obstacles when designing accurate localization algorithms.
\section{Conclusions}\label{sec:Conclusions}
In this work, we proposed an outlier-resistant fusion algorithm for multi-static positioning, addressing the challenges outlier measurements pose. The proposed algorithm employs the Cauchy loss function to reduce the impact of the outliers that arise from the multipath environment and the inherent characteristics of extended targets. The algorithm uses a two-step approach that combines AoA-based initialization with GD to estimate positions. First, we employed the VMF likelihood function to calculate the initial position. Then, we applied GD to refine the position estimation.
Simulation results demonstrated that our method outperforms conventional benchmarks in positioning accuracy, including $\ell_2$, IRLS, and Huber loss functions. Additionally, the results are verified using Kalman filter tracking, demonstrating performance in both an empty room and cluttered with obstacles. 

\section*{Acknowledgment}
This work was funded by the Federal Ministry of Research, Technology and Space Germany within the projects ”KOMSENS-6G” under grant 16KISK128 and "SENSATION" under grant 16KIS2529.

\bibliographystyle{IEEEtran}
\bibliography{references.bib}

\end{document}